\documentclass[structabstract]{aa}  
\usepackage{graphicx}
\usepackage{longtable}
\usepackage{txfonts}
\usepackage{natbib}
\bibpunct{(}{)}{;}{a}{}{,}
\begin{document}
   \title{Evidence for enhanced chromospheric Ca II H \& K emission in
stars with
close-in extrasolar planets}


   \author{T. Krej\v{c}ov\'{a}
          \inst{1}
          \and
          J. Budaj\inst{2}
          }

   \institute{Department of Theoretical Physics and Astrophysics, 
Masaryk University, Kotl\'{a}\v{r}sk\'{a} 2, 61137 Brno, Czech Republic\\
              \email{terak@physics.muni.cz}
         \and
             Astronomical Institute, Slovak Academy of Sciences,
             05960 Tatransk\'{a} Lomnica, Slovak Republic\\
             \email{budaj@ta3.sk}
             }

   \date{Received September 15, 1996; accepted March 16, 1997}

 
  \abstract
   {The planet-star interaction is manifested in many ways. 
It was found out that a close-in exoplanet causes small but measurable 
variability in the cores of a few lines in the spectra of several stars
which corresponds to the orbital period of the exoplanet.
Stars with and without exoplanets may have different properties.}
   {The main goal of our study is to search for influence which exoplanets 
might have on atmospheres of their host stars. Unlike the previous
studies, we do not study changes in the spectrum of a host star or 
differences between stars with and without exoplanets.
We aim to study a large number of stars with exoplanets, current level of 
their chromospheric activity and look for a possible correlation with
the exoplanetary properties. 
}
   {
To analyse the chromospheric activity of stars we exploit our
own\thanks{2.2\,m ESO/MPG telescope, Program 085.C-0743(A)} 
and publicly available archival spectra\thanks{This research has made use of the
Keck 
Observatory Archive (KOA), which is operated by the W.\,M.\,Keck Observatory 
and the NASA Exoplanet Science Institute (NExScI), under contract with 
the National Aeronautics and Space Administration.}, 
measure the equivalent widths of the cores of Ca\,II\,H and K lines
and use them as 
a tracer of their activity. Subsequently, we search for their
dependence on 
the orbital parameters and mass of the exoplanet.
}
   {We found a statistically significant evidence that the equivalent
width of the Ca\,II\,K line emission and $\log R'_{\mathrm{HK}}$ activity
parameter of the host star varies with the semi-major axis and mass
of the exoplanet.
Stars with $T_{\mathrm{eff}}\leq 5500$\,K having exoplanets with semi-major 
axis $a\leq 0.15$\,AU ($P_{\mathrm{orb}} \leq 20$\,days) have a broad range of
Ca\,II\,K emissions and much stronger emission in general than stars
at similar temperatures but with higher values of semi-major axes.
Ca\,II\,K emission of cold stars
($T_{\mathrm{eff}} \leq 5500$\,K) with close-in exoplanets 
($a\leq 0.15$\,AU) is also more pronounced for more massive 
exoplanets. 
}
   {The overall level of the chromospheric activity of stars may be
affected by their close-in exoplanets. Stars with massive
close-in 
exoplanets may be more active.}


\maketitle
%
\section{Introduction}
There is a wide variety of interactions which may occur between a
close-in planet and its host star: evaporation of the planet \citep{Vidal03, Hubbard07}, 
precession of the periastron due to the general relativity, tides or other 
planets \citep{Jordan08}, synchronization and circularization of 
the planet's rotation and orbit, strong irradiation of the planet and
its effect on the planet radius \citep{Guillot02, Burrows07}, atmosphere
or stratosphere \citep{Hubeny03, Burrows08, Fortney08}.
However, not only the planet suffers from this interaction.  
Hot Jupiters also rise strong tides on their parent stars and
some of the stars can become synchronized as well. 
Dissipation of the tides can lead to an extra heating of the stellar 
atmosphere. Magnetic field of the exoplanet may interact with the stellar 
magnetic structures in a very complicated manner which is not very well 
understood and is a subject of recent studies \citep{Cuntz00, 
Rubenstein00, Ip04, Lanza08}. Such type of interaction might be observed mainly
in the chromospheres and coronae of the parent stars.

The planet-star interaction is currently examined across the whole energy 
spectrum beginning from X-rays to the radio waves and from the ground as well as from
the
space. A planet-induced X-ray emission in the system HD179949 was observed 
by \citet{Saar08}. On the other hand, the search for correlation
between X-ray luminosity and exoplanetary parameters 
\citep{Kashyap08, Poppenhaeger11} revealed quite contradictory results. Based on
the observations in the optical region, \citet{Shkolnik05,Shkolnik08} 
discovered variability in the cores of 
Ca\,II\,H\,\&\,K, H$\alpha$ and Ca\,II\,IR triplet in a few exoplanet host stars
induced by the
exoplanet.

Ca\,II\,H\,\&\,K lines (3933.7 and 3968.5 \AA) are one of the best indicators 
of stellar activity observable from the ground \citep{Wilson68}.
These lines are very strong and their core is formed in the chromosphere
of the star. For quantitative assessment of chromospheric activity of
stars of different spectral types, the chromosphere emission ratio 
$\log R'_{\mathrm{HK}}$ (the ratio of the emission from the chromosphere 
in the cores of the Ca\,II\,H\,\&\,K lines to the total bolometric emission 
of the star) was introduced by \citet{Noyes84}.
Large $\log R'_{\mathrm{HK}}$ values mean higher activity.

\citet{Knutson10} found a correlation between 
the presence of the exoplanetary stratosphere and the Ca\,II\,H\,\&\,K 
activity index $\log R'_{\mathrm{HK}}$ of the star. Consequently, \citet{Hartman10} found a correlation between 
the surface gravity of Hot Jupiters and the stellar activity.

Recently, \citet{Canto_Martins11} analysed the sample of stars with and without
extrasolar
planets. They searched for 
a correlation 
between planetary parameters and the $\log R'_{\mathrm{HK}}$ parameter but 
didn't reveal any convincing proof for enhanced planet-induced activity in 
the chromosphere of the stars.
On the other hand, \citet{Gonzalez11} claims that stars with exoplanets
have smaller $v\sin i$ and $\log R'_{\mathrm{HK}}$ values (i.e. less activity)
than stars without exoplanets.

In this paper, we search for possible correlations between chromospheric
activity of the star and properties of exoplanets.
The equivalent width (hereafter EQW) of Ca\,II\,K line emission is
used as 
an indicator of the level of chromospheric activity of the parent star.

\begin{figure}
\centering
\includegraphics[width=8.7cm]{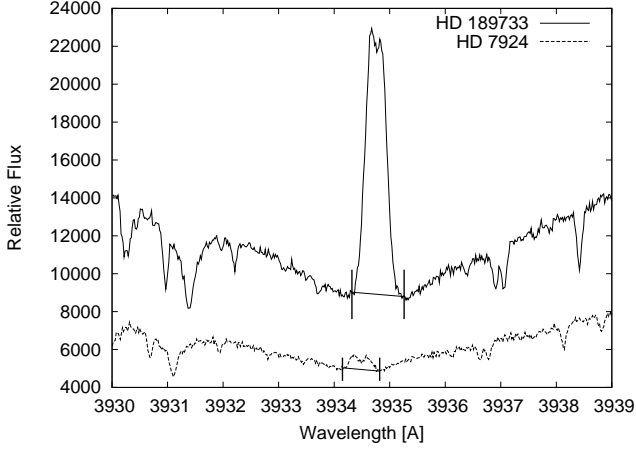}
\caption{Illustration of the Ca\,II\,K line in two different stars.
We measured only the equivalent width of core of the line with the
central
reversal. By definition, EQW is negative for emission and positive for
absorption.
}
\label{caiik}
\end{figure}

\section{Observation \& Data}
Stellar spectra used in our analysis originate from two different sources.
The main source is
the publicly available Keck HIRES spectrograph archive. These spectra have a typical resolution of up to 85\,000. 
We carefully selected only those spectra with the signal-to-noise ratio high
enough to measure the precise EQW of Ca\,II\,K emission.

These data were accompanied by our own observations
of several stars (HD\,179949, HD\,212301, HD\,149143 and Wasp-18) with 
close-in exoplanet. We obtained these data with FEROS spectrograph mounted on 
2.2\,m ESO/MPG telescope (night 18/19.9.2010). Spectra were reduced with standard procedure using IRAF
package\footnote{http://iraf.noao.edu/}. These data are marked in all
plots as red squares. We measured the EQW (in $\AA$) of 
the central reversal in the core of Ca\,II\,K from all spectra
using IRAF (see Figure \ref{caiik}). This plot illustrates a
placement of the pseudocontinuum in our measurements. Advantage of
using such simple EQW measurements is that they are defined on a short 
spectral interval which is about 1\,$\AA$ wide. Consequently, they are not 
very sensitive to various calibrations (continuum rectification, 
blaze function removal) inherent in the echelle spectroscopy.
For comparison, $\log R'_{\mathrm{HK}}$ index relies on the information
from four spectral channels covering about 100\,$\AA$ wide interval.
If extracted from echelle spectra it is much more sensitive to a proper 
flux calibration and subject to added uncertainties. 
 
The sample of the data contains 206+4 stars with extrasolar planets in the
temperature range from approx. 4\,500 to 6\,600\,K. The semi-major axes of 
the exoplanets lie in the interval 0.016--5.15\,AU. Table
\ref{t1} lists the parameters of exoplanetary systems used in the
study.

For comparison with our EQW values, we also used the parameter 
$\log R'_{\mathrm{HK}}$ taken from the work of \citet{Wright04}, \citet{Knutson10}
and
\citet{Isaacson10}.

\begin{figure}
\centering
\includegraphics[width=8.7cm]{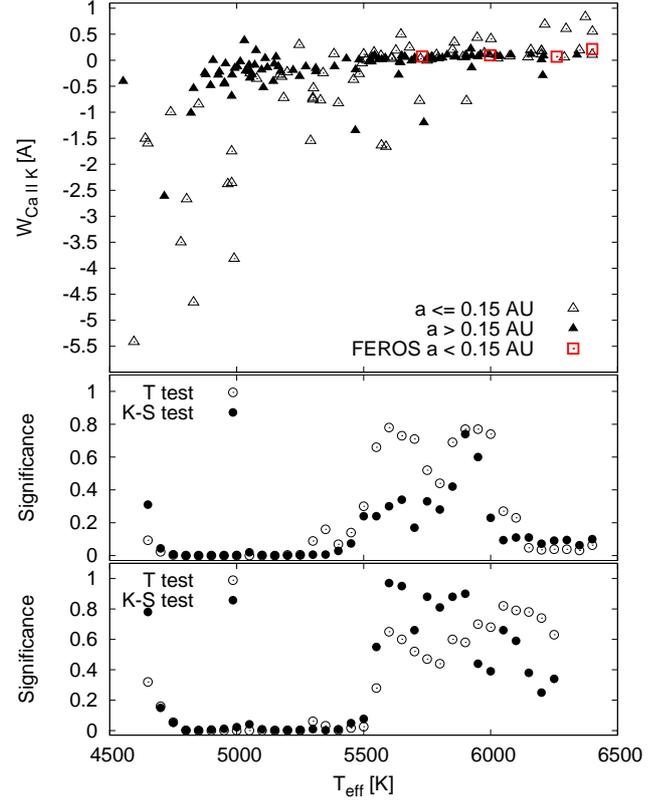}
\caption{{\it Top}: Dependence of the equivalent width of the
Ca\,II\,K 
emission on temperature of the parent star. Emission and
chromospheric 
activity increase with decreasing temperature. 
Cooler stars ($T_{\mathrm{eff}}\leq 5\,500$\,K) with close-in planets
are more 
active than stars with more distant planets.
Empty triangles are exoplanetary systems with $a\leq 0.15$\,AU, full 
triangles are systems with $a>0.15$\,AU and red squares are our data 
from FEROS. {\it Middle}: Statistical Student's \textit{t}-test (empty
circles) and 
Kolmogorov-Smirnov test (full circles) show whether the two
distributions 
are the same. Low probability values for $T_{\mathrm{eff}}\leq
5\,500$\,K mean 
a significant evidence that the chromospheric activity of stars with
close-in 
planets is different from that of stars with distant planets. 
{\it Bottom}: The same statistical tests performed on systems 
discovered by RV technique only.}
\label{Teff_W}
\end{figure}

\section{Statistical analysis and results}
\begin{figure}
\centering
\includegraphics[width=8.7cm]{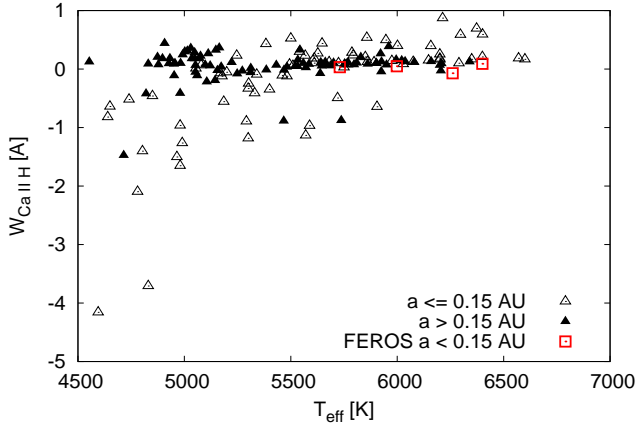}
\caption{Dependence of the equivalent width of Ca\,II\,H emission on 
the temperature of the star. Empty triangles are exoplanetary systems with
$a\leq 0.15$\,AU, full 
triangles are systems with $a>0.15$\,AU and red squares are our data 
from FEROS.}
\label{Tef_H}
\end{figure}

\begin{figure}
\centering
\includegraphics[width=8.7cm]{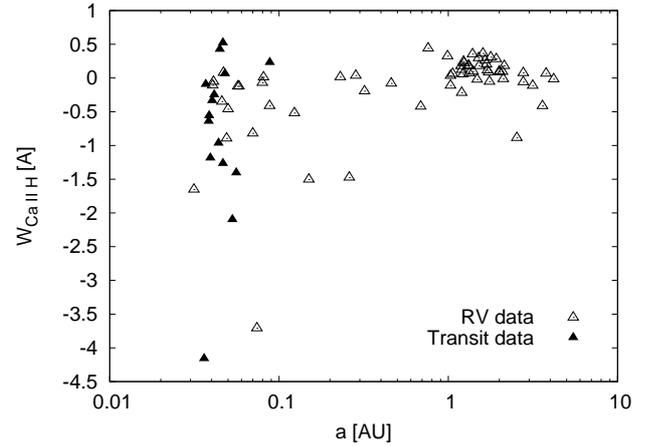}
\caption{Dependence of the equivalent width of Ca\,II\,H emission on 
the semi-major axis. Empty
triangles are exoplanetary systems 
discovered by RV technique, full triangles are systems discovered by transit
method.}
\label{a_H}
\end{figure}

We will start by exploring the dependence of the EQW of
Ca\,II\,K emission on the effective temperature of the star.
This is illustrated in the top panel of Figure \ref{Teff_W}\footnote{For comparison we also measured the Ca\,II\,H emission (figure \ref{Tef_H} and
\ref{a_H}). These are $T_{\mathrm{eff}}$ vs. $W_{\mathrm{Ca
II H}}$ and semi-major axis vs. $W_{\mathrm{Ca II H}}$ plots.}. One can
see that the EQW decreases i.e. core emission increases towards
lower temperatures.
Part of the reason for this behaviour is that the
photospheric flux at the core of the Ca II K line is lower for cooler
stars than for hotter stars.
At the same time, the data points which stand for stars with
$T_{\rm eff}> 5\,500$\,K show only
a narrow spread of Ca\,II\,K EQWs while cooler stars have
a broad range of these values. This means that we should focus on the
cooler stars.

\begin{figure}
\centering
\includegraphics[width=8.7cm]{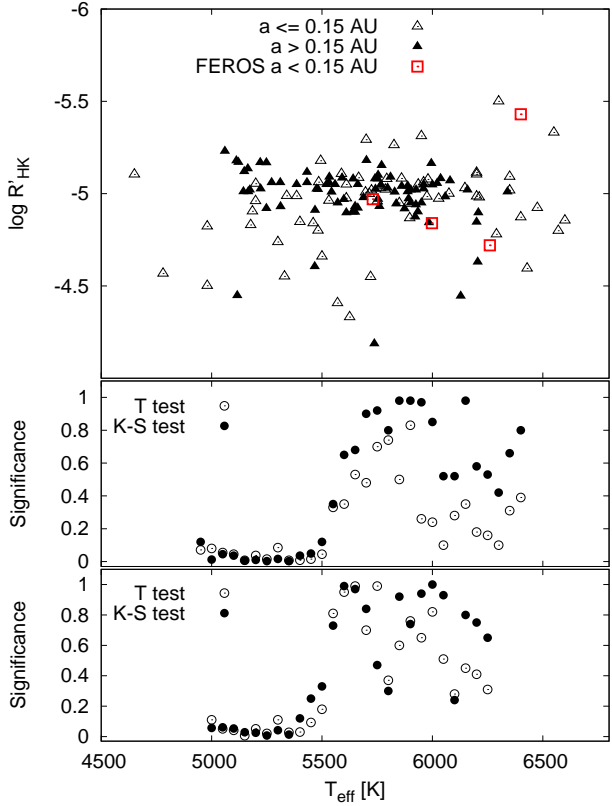}
\caption{{\it Top}: Dependence of the activity index $\log R'_{\mathrm{HK}}$ on
temperature of the
parent star. Emission and chromospheric activity 
increase with decreasing temperature. 
Cooler stars ($T_{\mathrm{eff}}\leq 5\,500$\,K) with close-in planets are more active
than
stars with more distant planets.
Empty triangles are
exoplanetary systems with $a\leq 0.15$\,AU, full triangles are systems with
$a>0.15$\,AU and red squares are our data from FEROS. 
{\it Middle}: Statistical Student's \textit{t}-test (empty circles) and 
Kolmogorov-Smirnov test (full circles) show whether the two distributions are 
the same. Low probability values for $T_{\mathrm{eff}}\leq 5\,500$\,K mean a
significant evidence 
that the chromospheric activity of stars with close-in planets is different 
from that of stars with distant planets. {\it Bottom}: The same 
statistical tests performed on systems discovered by RV technique only.}
\label{Rhk_teff}
\end{figure}

\begin{figure}[b]
\centering
\includegraphics[width=8.7cm]{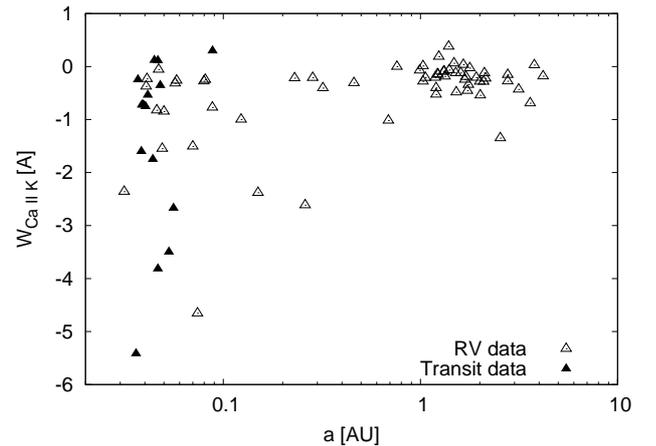}
\caption{Dependence of the equivalent width of Ca\,II\,K emission on 
the semi-major axis. Included are only systems with $T_{\mathrm{eff}}\leq 5\,500$\,K.
Empty
triangles are exoplanetary systems 
discovered by RV technique, full triangles are systems discovered by transit method.
Cold stars with
close-in planets ($a \leq 0.15$\,AU) 
have higher scatter and a generally higher Ca\,II\,K emission than cold stars 
with distant planets. This does not apply to hotter stars.}
\label{Perioda_W}
\end{figure}

In the next step, we will distinguish between the close-in and distant 
exoplanets. This is also shown in Figure \ref{Teff_W} where planets with
semi-major axis shorter/longer than 0.15\,AU have different symbols.
One can see that stars with close-in planets clearly tend to have higher 
Ca\,II\,K emission (lower EQWs) than stars with distant planets.
To verify whether this finding is statistically significant,
we performed two statistical tests on these two data samples (close-in vs
distant planets). The first one was the Student's \textit{t}-test which determines 
whether the means of these two samples are equal. 
The other test was Kolmogorov-Smirnov test which we used to determine 
whether the two groups originate from the same population.
We selected a running window which is 400\,K wide
and runs along the x-axis (temperature) with the step of 50\,K.
Consequently, we performed the statistical tests on the two samples of stars 
within the current window and plot the result versus the center of the
current window. The middle panel of Figure \ref{Teff_W}
shows 
the resulting
probability (as a function of temperature) that the two samples have 
the same mean or originate from the same population. 
Small value of probability means that the two samples
are different. It can be seen (figure \ref{Teff_W}), that the
difference between the stars 
with close-in and distant exoplanets is statistically significant for cooler 
stars with $T_{\mathrm{eff}} \leq 5\,500$\,K.

However, notice that while most of the distant exoplanets were
discovered by the radial velocity (RV) measurements, many of the close-in 
planets were discovered by transits. The two techniques may have very
different criteria for selection of the exoplanetary
candidates and especially the RV measurements concentrate 
on low activity stars. That is why in the
bottom panel of Figure
\ref{Teff_W}, we included only the stars with exoplanets 
discovered by the RV technique into the statistics. On this
reduced data sample we performed the same statistical tests as in the
previous case -- we chose the same size of the step and the running
window. Even if the transit data are excluded, the
tests 
show significant difference between the stars with close-in and distant 
exoplanets.

To verify the above mentioned trends, we also explored the
$\log R'_{\mathrm{HK}}$ parameter. This parameter does not show
the strong temperature dependence (Figure \ref{Rhk_teff}).
When we distinguish the stars with close-in exoplanets ($a \leq 0.15$\,AU) 
from stars with distant exoplanets ($a>0.15$\,AU) using different symbols we observe 
the same tendency as before. Namely, stars with close-in planets show
wider range of $\log R'_{\mathrm{HK}}$ values than stars with distant
planets. Cooler stars ($T_{\mathrm{eff}} \leq 5\,500$\,K)
with close-in planets have higher values of $\log R'_{\mathrm{HK}}$
and thus higher chromospheric activity.
The difference between stars with close-in and distant planets is
statistically significant for $T_{\mathrm{eff}} \leq 5\,500$\,K
which is illustrated by the Student's \textit{t}-test and  Kolmogorov-Smirnov
test in the middle panel of Figure \ref{Rhk_teff}. 
The bottom panel of the figure depicts the same tests applied to 
the systems discovered by the RV technique only and indicates
that the difference is statistically significant even though the sample 
consists of stars with planets detected by a single technique.
In this figure, we used 
the same kind of analysis within a running window as before.

\begin{figure}
\centering
\includegraphics[width=8.7cm]{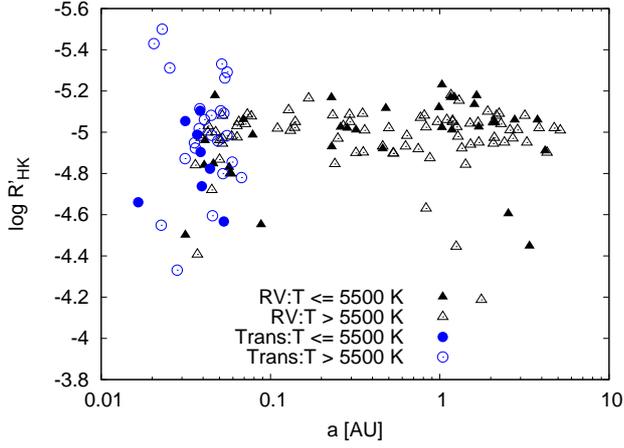}
\caption{Dependence of the activity index $\log R'_{\mathrm{HK}}$ on 
the semi-major axis. Triangles are exoplanetary systems 
discovered by RV technique, circles are systems discovered by transit method.}
\label{Rhk_porb}
\end{figure}

It appears that the semi-major axis of the innermost planet around a
star is in some way connected with the chromospheric activity of the star.
So, in the next step, we explore
the dependence of the activity on the semi-major axis~$a$.
This dependence is illustrated in Figure \ref{Perioda_W}
which displays results we measured --
EQW of Ca\,II\,K emission as a function of the semi-major
axis.
Following our findings above, we selected only systems with 
$T_{\mathrm{eff}} \leq 5\,500$\,K. 
One can clearly see two distinctive populations there.
Stars with close-in exoplanets with $a\leq 0.15$\,AU have a broad range of 
Ca\,II\,K emission while stars with distant planets ($a> 0.15$\,AU) 
have a narrow range of small Ca\,II\,K emission. 
Again, we distinguish between stars with planets discovered by 
the RV and transit methods. Significant fraction of close-in planets was
discovered by the RV method. Apparently, some stars with close-in exoplanets (but not all
of them) have high
Ca\,II\,K emission. 
Unfortunately, this finding may be affected by selection biases (it is more difficult
to detect distant planets around active stars).

To verify the above mentioned behaviour, we also explored 
the $\log R'_{\mathrm{HK}}$ parameter as a function of the semi-major
axis.
This is illustrated in Figure \ref{Rhk_porb}. 
This figure also shows
a clear distinction between the stars with close-in planets
with semi-major axis less than 0.15\,AU and stars with distant planets.
Stars with close-in exoplanets generally have higher scatter in the 
$\log R'_{\mathrm{HK}}$ values than stars with distant planets.
This is mainly caused by hotter stars with transiting exoplanets.
Once we concentrate only on cold stars with planets
detected by the RV technique there might be a trend that stars with 
close-in planets have higher $\log R'_{\mathrm{HK}}$ values than stars 
with distant planets but it is not as pronounced as in EQW
measurements. At the same time, there is no strong correlation with the semi-major axis
for $a\leq 0.15$\,AU.

\begin{figure}[b] 
\centering
\includegraphics[width=8.7cm]{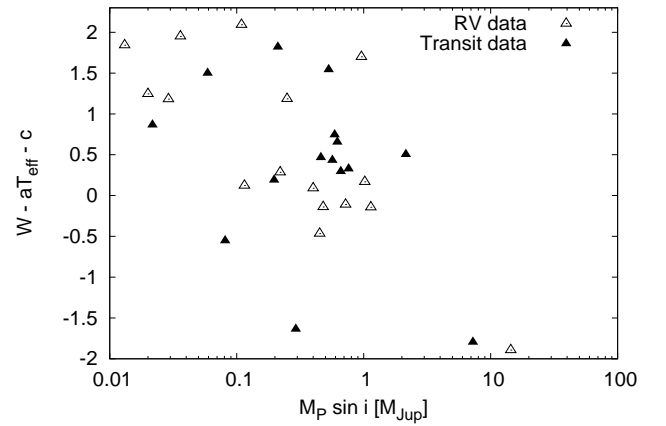}
\caption{Dependence of the modified equivalent width of Ca\,II\,K emission
on planet mass. Only stars with $T_{\mathrm{eff}}\leq 5\,500$\,K and 
$a\leq 0.15$\,AU 
are considered. Modified equivalent width is equivalent width
corrected for strong temperature dependence. By definition, the
EQW is negative for emission. Consequently, lower values mean a
higher emission and a higher chromospheric activity.
Chromospheric activity of stars with more massive 
planets is higher than in those with less massive planets.
Empty triangles - stars with planets detected by the RV method, full
triangles - transit method.}
\label{W_M}
\end{figure}

However, what causes the high range of values
of Ca\,II\,K emission for cold stars with close-in planets?
Is it due to its dependence on time? If yes, on what timescales?
On the timescale of the planet orbital period, stellar activity cycle,
age of the star or something else?
Is there any other parameter/process which affects the stellar chromosphere?
We explored whether it may be due to the eccentricity of the orbit 
but we did not find any convincing evidence for the eccentricity effect.
If there is a dependence of the Ca\,II\,K emission on 
the semi-major axis of the planet, there ought to be some dependence
on the mass $m_{\mathrm{p}}$ of the planet (or its magnetic field)
as well. This dependence would have to be
continuously reduced in case of sufficiently small planets.
Unfortunately, 
we know only $m_{\mathrm{p}}\sin i$ for most of the extrasolar
planets. 
Nevertheless, we select all stars with temperatures $T \leq 5\,500$\,K 
and with semi-major axis $a \leq 0.15$\,AU and 
fit EQWs of the Ca\,II\,K emission by the following
function:
\begin{equation}\label{eq1}
W(m_{\mathrm{p}},T_{\mathrm{eff}})=a T_{\mathrm{eff}} + b \log(m_{\mathrm{p}}\sin
i)+c
\end{equation} 
We found the following coefficients:
$a=3.65 \times 10^{-3}\pm 5.7 \times 10^{-4}, b = -0.392 \pm 0.097, c =
-20.4 \pm 2.9.$
The $a$ coefficient is significant beyond 6$\sigma$ and temperature 
dependence is thus very clear. However, the $b$ coefficient
is also significant beyond 4$\sigma$ and it indicates
the statistically significant correlation of the chromospheric
activity with the mass of the planet. We applied statistical \textit{F}-test to justify the usage of
additional parameter
(planet mass) in the above mentioned fitting procedure. The test shows that the
significance of the 3-parameter fit (equation (\ref{eq1})) over the
2-parameter fit $W(T_{\mathrm{eff}})= aT_{\mathrm{eff}} + c$ is 0.003, which is below the common 0.05 value (2$\sigma$).

The above mentioned results are illustrated in Figure \ref{W_M} where
we plot the modified equivalent width of the Ca\,II\,K emission
as a function of $m_{\mathrm{p}}\sin i$. Stars with planets detected
by RV and transit methods are marked with empty and full triangles respectively.
Modified equivalent width is EQW corrected for the strong temperature dependence,
namely $EQW-aT_{\mathrm{eff}}-c$.
Lower values mean larger emission and thus more massive planets
show more activity of the host star.
Unfortunately, this correlation might also be affected by 
the selection effect that it is more difficult to detect a less massive
planet around a more active star. 

If our findings based on the Ca\,II\,K emission are true, than
one can ask a question: how does this planet-star interaction work
and why does it operate up to $a=0.15$\,AU?
If it was caused by the tides of the planet, one would expect that 
the stellar activity would gradually decrease with $a$ which 
may not be the case. 
We suggest that it may be due
to the magnetic interaction which scales with the magnetic field of 
the planet. Magnetic field of the planet may be very sensitive to 
the rotation profile of the planet which is a subject to strong 
synchronization.
\citet{Bodenheimer11} suggested that exoplanets with semi-major axis 
$a \leq 0.15$\,AU are most probably synchronized.

\onllongtab{1}{
\begin{longtable}{lccccccc}
\caption{\label{t1}Parameters of exoplanetary systems. Data taken from
\texttt{http://www.exoplanet.eu/}. Systems below the line at the
bottom part of the table describe our
FEROS
data
sample.}\\
\hline\hline
Object&$M_{\mathrm{p}} \sin i$ [M$_{\mathrm{Jup}}$]&$P_{\mathrm{orb}}$
[days]&$a$\,[AU]&$T_{\mathrm{eff}}$\,[K]&W$_{\mathrm{Ca\,II\,K}}$
$[\AA]$&$\log
R'_{\mathrm{HK}}$&Detection method\\
\hline
WASP-19 b   	&   1.168   &      0.78884       &   0.01655 &  5500	&	-
	&	-4.66	 & T \\
Kepler-10 b  & 0.0143 &  0.837495  & 0.01684	& 5627 	&		0.1864&- &
T\\
WASP-12 b   	&   1.404   &      1.0914222        & 0.02293    & 6300	&	-
	&	-5.5	& T  \\  
HAT-P-23 b  & 2.09 &  1.212884 &0.0232	& 5905  	&	-0.7840 & - & T\\
TrES-3   	&   1.91   &     1.30618608   &   0.0226     &      5720 &     
-0.7795	&	-4.549	   & T \\
CoRoT-1 b   	&   1.03   &      1.5089557  &    0.0254 &      5950	&	-
	&	-5.312	  & T\\
CoRoT-2 b   	&   3.31   &      1.7429964  & 0.0281    &      5625 &	-	
&	-4.331	 & T   \\
WASP-3 b   	&   2.06   &     1.8468372    & 0.0313    &      6400 &      0.1073
	&	-4.872	 & T \\
HD 86081 b   	&   1.5    &     2.1375      &    0.039 &      6028 &      0.0371
	&	-4.973	& RV \\
HAT-P-32 b  & 0.941 &  2.150009  &  0.0344	&6001  &	0.4073& - & T\\
WASP-2 b   	&   0.847  &      2.15222144   &  0.03138   &      5150	&	-
	&	-5.054	  & T \\
HAT-P-7 b     &   1.8    &     2.2047298   &  0.0379   &      6350 &      0.1946     
	&	-5.018	  & T \\
HD 189733 b   &   1.138   &     2.21857312   &  0.03142   &      4980 &      -2.3605 
   	&	-4.501	  & RV \\
WASP-14 b   	&   7.341  &      2.2437661  &  0.036   &      6475	&	-
	&	-4.923	& T  \\
TrES-2   	&   1.253  &     2.470614    &  0.03556   &      5850 &      0.2015
   	&	-4.949	& T \\
WASP-1 b     	&   0.86   &     2.5199464     &  0.0382   &      6200 &      0.1785 
    	&	-5.114	& T  \\
HD 73256 b   	&   1.87   &     2.54858     &  0.037   &      5570 &      -1.6362   
 	&	-4.407	&RV  \\
XO-2 b   	&   0.62   &     2.615838    &  0.0369   &      5340 &      -0.2480  
  	&	-4.988	 & T \\
HAT-P-16 b  & 4.193  & 2.77596	&0.0413	 & 6158  	&	0.1982 & - & T\\
HAT-P-5 b    	&  1.06      &   2.788491   & 0.04075  &        5960	&	-
&-5.061 & T	 \\
HAT-P-20 b  & 7.246  & 2.875317  &	0.0361&  4595  	&	-5.4172	& - & T\\
HD 149026 b  	&    0.356   &    2.8758916    &    0.04288  &    6147    &    
0.0581	   &-5.030	&RV  \\
HAT-P-3 b     &    0.591   &    2.899703     & 0.03866     &    5185    &  -0.7236
&-4.904	& T  \\
HD 83443 b   	&    0.4     &    2.985625     &  0.0406    &    5460    &    
-0.3757   &-4.84	& RV  \\
HD 46375 b   	&    0.249   &    3.024        &  0.041    &    5199    &     -0.2325
   &-4.96	& RV  \\
TrES-1       	&  0.761      &  3.0300722      &  0.0393    &  5300      &    
-0.7222    &-4.738	& T  \\
HAT-P-27 b/WASP-40 b&  0.66 &  3.0395721& 0.0403 & 5300  &		-0.7542& - &
T\\
HAT-P-4 b   	&  0.68      &  3.0565114       & 0.0446    &   5860      &   0.0739 
   &-5.082	& T  \\
HAT-P-8 b   & 1.34    &    3.07632402         & 0.0449     &6200 &        0.1113  &
-4.985	       & T    \\
HD 187123 b    &0.52  &  3.0965828  &  0.0426	&5714    & - &                       
   -4.999 & RV\\
XO-3 b   	& 11.79   &     3.1915239      & 0.0454     &6429	&	-
&	-4.595	   & T      \\
HAT-P-22 b  & 2.147 &  3.21222   & 0.0414  &5302  	&	-0.5378 & - & T\\
HAT-P-12 b   	& 0.211   &    3.2130598       &   0.0384  & 4650 &        -1.6008  &
 	-5.104	  & T       \\
Kepler-4 b &  0.077  & 3.21346  &    0.0456 	&5857  	&	0.3420 &- & T\\
Kepler-6 b  & 0.669 &  3.234723  & 0.04567 	&5647  	&	0.4951&- & T\\
HAT-P-28 b &  0.626 &  3.257215  &   0.0434	&5680  	&	0.2449& - & T\\
HAT-P-24 b  & 0.685 &  3.35524  &  0.0465	&6373  &		0.8322 & - &
T\\
HD 88133 b   	& 0.22    &    3.416           &  0.047    &5494 &        -0.0570  & 
	-5.178 & RV	         \\
HAT-P-33 b  & 0.763  & 3.474474  &  0.0503	&6401  	&	0.5521&- & T\\
BD-10 3166 b  & 0.48    &    3.487           &  0.046    &5400 &        -0.8254	 &  
-4.847	   & RV      \\
Kepler-18 b  & 0.0217 &  3.504725   & 0.0447	&5383  	&	0.1184&- & T\\
Kepler-8 b &  0.603  & 3.52254   &  0.0483	&6213  	&	0.6870&- & T\\
HD 209458 b   & 0.714   &    3.52474859      & 0.04747     &6075 &         0.0744  & 
	-5.00	     & RV    \\
Kepler-5 b  & 2.114  & 3.54846   & 0.05064  &6297  	&	0.6038&- & T\\
TrES-4    	& 0.917   &    3.5539268        &   0.05084  & 6200 &         0.1783 
&  	-5.104	    & T     \\
HAT-P-25 b &  0.567 &  3.652836   & 0.0466	&5500  &		0.1131 & - &
T\\
WASP-11/HAT-P-10 b   	& 0.46    &    3.722469 &  0.0439   & 4980 &        -1.7503 &
-4.823	   & T        \\ 
WASP-17 b   	& 0.486    &     3.735438      &0.0515      &6550&		-
&	-5.331	  & T       \\
HD 219828 b   & 0.066   &    3.8335          &  0.052    &5891 &         0.0604  & 
-4.945	            & RV    \\
HAT-P-6 b   	& 1.057   &    3.853003        & 0.05235     &6570 &         0.1397 
&  -4.799	     & T    \\
HAT-P-9 b   	& 0.67    &     3.922814       & 0.053     &6350	&	-
&-5.092	     & T    \\
XO-1 b   	& 0.9     &    3.9415128       &   0.0488  &5750     &        -0.0056
&   -4.958	& T         \\
HAT-P-19 b &   0.292 &  4.008778  & 0.0466	&4990  	&	-3.8168 &- & T\\
HD 102195 b   & 0.45    &    4.113775        &  0.049   & 5291  &        -1.5490 &  
- & RV	                \\
HAT-P-21 b  & 4.063 &  4.124461  &0.0494  & 5588  	&	-1.6625 & - & T\\
XO-4 b   	& 1.72    &     4.1250823        &  0.0555   & 5700&		-
&-5.292	  & T       \\
HD 125612 c    &	0.058  &   4.1547 & 0.05  &5897     & - &             -4.867
& RV\\
XO-5 b   	& 1.077   &    4.1877537       & 0.0487     &5510 &         0.0093  
&	-	  & T       \\
61 Vir b   	&	0.016 &  4.215    & 0.050201	&5531                  &     
       -     & -4.962 & RV\\
51 Peg b   	& 0.468   &    4.23077         & 0.052     &5793 &         0.0432   &
-5.08	     &RV   \\
HAT-P-26 b & 0.059  & 4.234516   & 0.0479  &5079  	&	-0.3556	&-	& T\\
WASP-13 b   	& 0.485    &     4.353011        &  0.05379    &5826&		-
&	-5.263	      & T   \\
Kepler-12 b &  0.431 &  4.4379637  & 0.0556 &5947 & 	0.4331&- & T\\
HAT-P-1 b   	& 0.524   &    4.4652934       &  0.05535   & 5975 &         0.1351  
&	-4.984	  & T      \\
ups And b    &	0.69  &  4.617136  & 0.059  &6212        & - &                       
 -4.980 & RV\\
HAT-P-14 b   	& 2.2   &    4.627657       &  0.0594    &6600 &         0.1183   &
-4.855	     & T    \\
HD 156668 b   & 0.0131  &    4.646           & 0.05     &4850    &         -0.8485  &
-	    & RV           \\
HAT-P-11 b   	& 0.081   &    4.887804        & 0.053     &4780 &         -3.4991  &
-4.567	    & T     \\
HD 49674 b   	& 0.115   &    4.9437          & 0.058     &5482 &         -0.2634  &
-4.80	     & RV      \\
HD 109749 b   & 0.28    &    5.24            & 0.0635     &5610 &         0.0512
  & -4.975	    & RV   \\
HD 7924 b   	& 0.029   &    5.3978          &  0.057    &5177 &         -0.3167  &
-4.83 	   & RV    \\
HAT-P-18 b  & 0.197 &  5.508023  & 0.0559  &4803  &		-2.6726 &- & T\\
HAT-P-2 b    	& 8.74    &    5.6334729       &  0.0674    &6290 &         0.0552
  &-4.78	& T         \\
HD 1461 b   	& 0.0239  &    5.7727          & 0.063438     &5765 &         0.0179 
 & -5.03	     & RV      \\
HD 68988 b   	& 1.9     &     6.276          & 0.071    &	5767 	&
0.0495		&	-5.04 	 & RV \\
HD 168746 b   & 0.23    &    6.403           & 0.065     &5610 &	
0.0273		&	-5.05	 & RV \\
HD 102956 b  & 0.96  & 6.495 &0.081  	&5054  	&	-0.2472 & - & RV\\
HIP 14810 b    &	3.88  &  6.673855  & 0.0692  &5485    & - &                  
       -5.062 & RV\\
HD 185269 b    &	0.94 &   6.838  & 0.077 &5980       & - &                    
        -5.077 & RV\\
HD 217107 b    &	1.33  &  7.12689  &  0.073	&5666     & - &              
            -5.086 & RV\\
HIP 57274 b &  0.036  & 8.1352  &  0.07	&4640  	&	-1.5066	& -& RV\\
HD 162020 b   & 14.4   &    8.428198        & 0.074     &4830&		-4.6563	&
-& RV	  \\
HD 69830 b    &	0.033   & 8.667  & 0.0785 	&5385      &     -         &         
      -4.987 & RV\\
Kepler-19 b &  0.064 &  9.2869944  &  0.118	&5541  	&	0.1498&- & T\\
HD 97658 b &  0.02  & 9.4957  &  	0.0797	&5170  	&	-0.2781 &- & RV\\
HAT-P-17 b  & 0.53 &  10.338523 &  0.0882	&5246  	&	0.2963 &- & T\\
HD 130322 b   & 1.02    &     10.72         & 0.088     	&5330  	&
-0.7721	&	-4.552	 & RV \\
HAT-P-15 b  & 1.946 &  10.863502  & 0.0964 	&5568  	&	0.0856 &-  & T\\
HD 38529 b  &  	0.78  &  14.3104 &   0.131	&5697    &     -          &          
   -5.007 & RV\\
HD 179079 b   & 0.08    &     14.476         &       0.11	&5724 	&
0.0267		&	-5.018	& RV \\
HD 99492 b   	& 0.109   &     17.0431        &   0.1232    &4740	&
-0.9996	&	-	& RV\\
HD 190360 c    & 0.057 & 17.1&0.128	  & 5588     & - &                           
  -5.107 & RV\\
HD 16417 b & 0.069  &  17.24  & 0.14 	&5936        &     -	&        -5.050	& RV
\\
HD 33283 b   	& 0.33    &     18.179         &   0.168     &5995 	&
0.1073		&	-5.164	&RV\\	
HD 195019 b   & 3.7   &     18.20163       &   0.1388    &5787 	&	0.1036	
&	-5.022	&RV\\	
HD 192263 b   & 0.72    &     24.348         &  0.15  	&4965	&	-2.3811	&
-	& RV\\
HD 224693 b   & 0.71    &     26.73          &  0.233    & 6037  	&
0.0726		&	-5.082	&RV\\	
HD 43691 b   	& 2.49    &     36.96          &  0.24     &6200  	&
0.0103		&	-4.846	&RV\\	
HD 11964 c   &  0.079 & 37.91  & 0.229   &5248    & -  &                          
-5.168 & RV\\
HD 45652 b   	& 0.47    &     43.6           &  0.23     &5312 	&
-0.2185	&	-4.930	&RV\\	
HD 107148 b   & 0.21    &     48.056         & 0.269      &5797  	&
0.0290		&	-5.03	&RV\\	
HD 90156 b   & 	0.057  &  49.77  &0.25   &5599  &      -         &                 
-4.969 & RV\\
HD 74156 b   &	1.88  &  51.65  & 0.294 &5960     & - &                              
-5.050 & RV\\
HD 37605 b  &  	2.84  &  55.23  & 0.26 &5475    & - &                              
-5.025 & RV\\
HD 168443 b    &	7.659  &  58.11247  & 0.2931  &5591  & - &                   
      -5.085 & RV\\
HD 85512 b  & 0.011  & 58.43  &  0.26	&4715  	&	-2.6150& - & RV\\
HD 3651 b   	& 0.2     &     62.23          &  0.284     &5173	&
-0.2119	&	-5.02	&RV\\	
HD 178911 B b    &	6.292  &  71.487 &  0.32		&5650   & - &        
                 -4.900 & RV\\
Gl 785 b    &	0.053 &   74.72  &0.32  	& 5144         & -0.4055&            
             -5.011 & RV\\
HD 163607 b    &	0.77  &  75.229   &  0.36	&5543       & 0.0947 &       
                   -5.009 & RV\\
HD 16141 b   	& 0.215    &     75.82          & 0.35      &5533	&
0.0657		&	-5.09	&RV\\	
HD 114762 b  & 10.98  &  83.9151  &  0.353	& 5934        & - &                  
 -4.902 & RV\\
HD 80606 b   	& 3.94    &     111.43637      & 0.449      &5645 	&
-0.0441	&	-	& RV\\	
70 Vir b & 6.6	& 116.67   &   0.48	&5432        &           -	&            
     -5.116	& RV \\
HD 216770 b   & 0.65    &     118.45         &   0.46    &5248 	&	-0.3098	&
-4.92	&RV\\	
HD 52265 b   	& 1.05    &     119.6          &  0.5    	&6159  	&
0.1133		&	-5.02	&RV\\
HD 102365 b   & 	0.05  &  122.1   &   0.46	&5650         & 0.0863 &     
                   -4.931& RV\\
HD 231701 b   & 1.08    &     141.6          &     0.53 	&6208 	&
0.0829		&	-4.897	&RV\\	
HD 37124 b	&	0.675 &	154.378	&0.53364	&5610      & -    &          
                 -4.897 & RV\\
HD 11506 c & 0.82  & 170.46  &  0.639  	&6058        &  -  &                         
  -4.983 & RV\\
HD 5891 b  & 7.6  & 177.11  &  0.76	&4907 	&	-0.0017 & - & RV\\
HD 155358 b    &	0.89  &  195  &0.628   &5760       & - &                     
    -4.931 & RV\\
HD 82943 c    &	2.01  &  219   & 0.746  &5874        & - &                           
 -4.918 & RV\\
HD 218566 b  & 0.21  & 225.7  & 0.6873 	&4820  	&	-1.0175&- & RV\\
HD 8574 b   	& 2.11    &     227.55         &  0.77     &6080  	&
0.1015		&	-5.07& RV\\	
HD 134987 b    &	1.59  &  258.19  & 0.81   &5740        & -&                  
      -5.081& RV\\
HD 40979 b  	& 3.28    &     263.1          &  0.83     &6205 	&
-0.2931	&	-4.63 	&RV\\	
HD 12661 b  &  2.3  &  263.6  &0.83    &5742   & - &                               
-5.024 & RV\\
HD 164509 b   & 	0.48  &  282.4   &  0.875   &5922       & 0.2167 &           
                -4.874 &RV\\
HD 175541 b   & 0.61    &     297.3          &1.03       &5060 	&	-0.2814	&
-5.23	&RV\\	
HD 92788 b   	& 3.86    &     325.81         & 0.97      &5559 	&
0.0118		&	-5.05	&RV\\	
HD 33142 b  & 1.3  & 326.6  & 1.06	& 5052  	&	-0.2125 &- & RV\\
HD 192699 b    &	2.5  &  351.5  & 1.16	&5220    &         -           &     
     -5.169 & RV\\
HD 4313 b  & 2.3 &  356  &  	1.19	&5035  	&		-0.2078&- & RV\\
HD 96063 b    &	0.9  &  361.1  &   0.99	&5148   &     -0.0725 &                      
      -5.120 & RV\\
HD 212771 b    &	2.3  &  373.3   &  1.22 	&5121       & -0.1416 &      
                     -5.170 & RV\\
alf Ari b &  1.8 &  380.8  &  1.2	&4553  &		-0.4053 & - & RV\\
HD 28185 b   	& 5.7     &     383            &1.03       &5482 	&
0.0092		&	-5.023	&RV\\	
HD 28678 b  & 1.7  & 387.1  &  1.24		&5076  	&	0.1876 & - & RV\\
HD 75898 b   	& 2.51    &     418.2          &    1.19   	&6021  	&
0.0944		&	-5.056	&RV\\	
HD 4203 b   	& 2.07    &     431.88         &   1.164    &5701 	& 
0.0444		&	-5.18	&RV\\	
HD 1502 b &  3.1  & 431.8  &  1.31		&5049  	&	-0.0948& - & RV\\
HD 98219 b &  1.8  & 436.9  &   1.23		&4992  	&	-0.1541 &- & RV\\
HD 99109 b   	& 0.502   &     439.3          &    1.105    &5272	&
-0.1254	&	-5.06	&RV\\	
HD 210277 b   & 1.23    &     442.1          &   1.1      &5532	&	-0.0307	&
-5.06	&RV\\
HD 108863 b &  2.6 &  443.4  &  1.4			&4956  	&	-0.0677 &-&
RV\\
24 Sex b  & 1.99  & 452.8 &  	1.333		&5098  	&	-0.186 & -& RV\\
HD 188015 b   & 1.26    &     456.46         &  1.19     &5520	&	-0.0296	&
-5.05	&RV\\	
HD 136418 b  & 2 &  464.3 &  1.32			&5071  	&	-0.0939	& -&
RV\\
HD 31253 b  &  	0.5  &  466  &  1.26		&5960    & 0.1123 &                  
                 -5.026 &RV\\
HD 180902 b &  1.6  & 479  &  	1.39		&5030  	&	0.3773&- & RV\\
HD 96167 b   	& 0.68    &     498.9          &  1.3    & 5770 	&
0.0629		&	-5.153	&RV\\	
HD 20367 b    &	1.07  &  500  &  1.25	&6128    &        -              &           
-4.445 & RV\\
HD 114783 b   & 1.0    &     501            &  1.2    	&5105	&	-0.5261	&
-	&RV\\	
HD 95089 b  & 1.2  & 507   &  1.51	&5002  	&	-0.1224 &- & RV\\
HD 158038 b  & 1.8  & 521  &  1.52 	&4897  	&	-0.4858 &- & RV\\
HD 192310 c    &	0.075  &  525.8   &  1.18	&5166    & - &               
           -5.011 & RV\\
HD 4113 b   	& 1.56    &     526.62         &  1.28    & 5688	&
-0.0128	&	-4.979	&RV\\	
HD 19994 b   & 	1.68  &  535.7   & 	1.42	&5984 &          -               &   
 -4.843 & RV\\
HD 222582 b   & 7.75    &     572.38         &  1.35     &5662	&	0.0567	
&	-4.922	&RV\\	
HD 206610 b  & 2.2 &  610  & 1.68 	&4874  	&	-0.2457&- & RV\\
HD 200964 b    &	1.85  &  613.8  & 1.601  &5164          & -0.1220 &          
              -5.135 &RV\\
HD 183263 b    &	3.67  &  626.5  &  1.51 &5888  & - &                         
      -5.042 & RV\\
HD 141937 b   & 9.7     &     653.22         &   1.52    &5925 	&	-0.1438	&
-4.94	&RV\\	
HD 181342 b  & 3.3  & 663   &1.78   &5014  	&	-0.0290&- & RV\\
HD 116029 b &  2.1  & 670.2 & 1.73  &4951  	&	-0.4563&- & RV\\
HD 5319 b   	& 1.94    &     675            & 1.75      &5052	&
-0.3424	&	-	&RV\\		
HD 152581 b  & 1.5 &  689  & 1.48	& 5155  	&	0.0610&- & RV\\
HD 38801 b  & 	10.7  &  696.3  & 1.7 	&5222      &     -0.1931        &            
    -5.026 & RV\\
HD 82886 b    &	1.3   & 705  &  1.65	&5112    & 0.0341 &                          
        -5.178 & RV\\
HD 18742 b  & 2.7  & 772 &  	1.92	&5048 		&	-0.2086 & - & RV\\
HD 102329 b  & 5.9  & 778.1  &  2.01 		&4830 	&	-0.5418 &- & RV\\
16 Cyg B b  &	1.68  &  799.5   &	1.68	&5766       & - &                 
-5.046 & RV\\
HD 4208 b   	& 0.8     &     829            &  1.7     &5571  	&
0.0141		&	-4.95	&RV\\	
HD 70573 b   	& 6.1     &     851.8          &   1.76   	&5737 	&
-1.2001	&	-4.187	&RV\\	
HD 99706 b  & 1.4  & 868  &  2.14		&4932  	&	-0.2245 &- & RV\\
HD 131496 b  & 2.2 &  883  &  2.09		&4927  	&	-0.2836& - & RV\\
HD 45350 b   	& 1.79    &     890.76         & 1.92      &5754 	&
0.0338		&	-5.10	&RV\\	
HD 30856 b &  1.8  & 912 &  	2	&4982  		&	-0.2822 &- & RV\\
HD 16175 b  &  4.4  &  990   &  2.1		&6000     & - &                      
       -5.048 & RV\\
HD 34445 b    &0.79  &  1049  &  2.07		&5836       & - &                    
         -4.945 & RV\\
HD 10697 b   	& 6.38    &     1076.4         & 2.16      &5641	&
-0.0343	&	-5.08	&RV\\	
47 Uma b   	&	2.53 &   1078    &2.1		&5892        &             - 
           &-4.973 & RV\\
HD 114729 b   & 0.84    &     1135           &2.08       &5662	&	0.0658	
&	-5.05	&RV\\	
HD 170469 b   & 0.67    &     1145           & 2.24     	&5810  	&
0.0452		&	-5.09	&RV\\	
HD 164922 b   & 0.36    &     1155           &  2.11    & 5385	&	-0.1236	&
-5.05	&RV\\	
HD 30562 b   	& 1.29    &     1157           &  2.3     &5861  	&
0.0667		&	-5.04	&RV\\	
HD 126614 b  & 0.38  & 1244  & 2.35  &5585  	&	0.0434 &- & RV\\
HD 50554 b   	& 5.16    &     1293          &  2.41     &5902 	&
0.0535		&	-4.95	&RV\\	
HD 142245 b &  1.9  & 1299  &  2.77		&4878  	&	-0.2781&- & RV\\
HD 196885 A b & 2.98    &     1326           &  2.6     &6340  	&	0.1021	
&	-5.01	&RV\\	
HD 171238 b   & 2.6     &     1523           &  2.54    	&5467	&
-1.3506	&	-4.605	&RV\\	
HD 23596 b    &	8.1  &  1565  &  2.88		&5888       & - &                    
         -5.011 & RV\\
HD 106252 b   & 7.56    &     1600           &  2.7     &5754  	&	0.0617	
&	-4.97	&RV\\	
14 Her b   	& 4.64    &     1773.4         &  2.77     &5311	&
-0.1555	&	-5.06	&RV\\	
HD 73534 b   	& 1.15    &     1800           & 3.15      &4952	&
-0.4324	&	-	&RV\\	
HD 66428 b   	& 2.82    &     1973           &  3.18     &5752  	&
0.0465		&	-5.08	&RV\\	
HD 89307 b   	& 1.78    &     2157           &  3.27     &5950  	&
0.0688		&	-4.95	&RV\\	
HD 50499 b   	& 1.71    &     2482.7         & 3.86      &5902  	&
0.1008		&	-5.02	&RV\\	
eps Eri b   &	1.55  &  2502  &   3.39 		&5116    &       - &         
            -4.448 & RV\\
HD 117207 b   & 2.06    &     2627.08        &  3.78     &5432	&	0.0276	
&	-5.06	&RV\\	
HD 87883 b   	& 1.78    &     2754           & 3.6      &4980	&	-0.6913	&
-	&RV\\	
HD 106270 b  &  	11  &  2890   &  4.3 		&5638        & -0.2846 &     
                      -4.901 & RV\\
HD 154345 b   & 0.947   &     3340           &  4.19      &5468	&	-0.1824	&
-4.91	&RV\\	
HD 72659 b   	& 3.15    &     3658           & 4.74 		&5926  	&
0.0844		&	-5.02	& RV\\
HD 13931 b   & 1.88  &  4218   &  	5.15	&5829      & 0.1010 &                
              -5.009 & RV\\
\hline		
HD 149143 b	     & 1.36	&4.088			&0.052	&5730	&0.0716	
&	-4.97	& RV\\                    
HD 179949 b	     & 0.95	&3.0925		& 0.045	&6260	&0.0678		&
-4.72	& RV\\      
HD 212301 b	     & 0.45	&2.245715	&0.036	&5998	&0.0966		&
-4.84	& RV\\             
WASP-18 b	     & 10.43	&0.9414518	&0.02047	&6400	&0.2145	
&	-5.43	& T \\             
             
\hline                                  
\end{longtable}
}

\section{Conclusions}

We have found a statistically significant evidence that EQW 
of the Ca\,II\,K emission in the spectra of planet host stars as well
as 
their $\log R'_{\mathrm{HK}}$ activity index depend on the semi-major axis 
of the exoplanet.
Stars with close-in planets ($a \leq 0.15$\,AU)
have a generally higher Ca\,II\,K emission than stars with more distant planets.
This means that a close-in planet may affect the level of the chromospheric
activity of its host star and might heat the chromosphere of the star.
This process operates up to the orbital period of about 20 days. Moreover, we have found a statistically significant evidence that the Ca\,II\,K
emission of the host star (for $T_{\mathrm{eff}}\leq 5\,500$\,K and 
$a\leq 0.15$\,AU) increases with the mass of the planet.
The above mentioned trends may be affected by selection 
effects and should be revisited when less biased sample of stars
with planets becomes available.

\begin{acknowledgements}
We thank anonymous referee for important comments and suggestions
on the manuscript.
This work has been supported by grant GA \v{C}R GD205/08/H005,
Student Project Grant at MU MUNI/A/0968/2009, the National scholarship 
programme of Slovak Republic, and by VEGA 2/0094/11, VEGA 2/0078/10
and VEGA 2/0074/09. 
We want to thank Tom\'a\v{s} Henych for fruitful discussion and Mark\'{e}ta
Hyne\v{s}ov\'{a} for language correction.
\end{acknowledgements}

\bibliographystyle{aa}
\bibliography{ekvivalent}

\end{document}